\documentclass[
showpacs
]{revtex4-1}
\usepackage{amsthm}
\theoremstyle{plain}
\newtheorem{theorem}{Theorem}
\newtheorem*{theorem*}{Theorem}

\newtheorem*{definition*}{Definition}
\newtheorem{lemma}[theorem]{Lemma}
\newtheorem*{lemma*}{Lemma}

\newtheorem{remark}[theorem]{Remark}

\usepackage{physics}

\usepackage{braket}
\usepackage{delarray}
\usepackage{here}

\usepackage{amsmath}
\usepackage{txfonts}

\usepackage[dvipdfmx]{graphicx}
\usepackage{bm}
\usepackage{color}
\usepackage{ulem}

\newcommand{\be}{\begin{eqnarray}}
\newcommand{\ee}{\end{eqnarray}}
\newcommand{\ba}{\begin{array}}
\newcommand{\ea}{\end{array}}
\newcommand{\bmat}{\left(\begin{array}}
\newcommand{\emat}{\end{array}\right)}
\newcommand{\no}{\nonumber}

\newcommand{\vect}[1]{\mbox{\boldmath $#1$}}
\newcommand{\svect}[1]{\mbox{\boldmath {\scriptsize${#1}$}}}

\begin{document}
\title{Mean-field theory is exact for Ising spin glass models with Kac potential in non-additive limit on Nishimori line}
\author{Manaka Okuyama$^1$}
\author{Masayuki Ohzeki$^{1,2,3}$}
\affiliation{$^1$Graduate School of Information Sciences, Tohoku University, Sendai 980-8579, Japan}
\affiliation{$^2$Institute of Innovative Research, Tokyo Institute of Technology, Yokohama 226-8503, Japan}
\affiliation{$^3$Sigma-i Co., Ltd., Tokyo 108-0075, Japan} 

\begin{abstract} 
Recently, Mori [Phys. Rev. E \textbf{84}, 031128 (2011)] has conjectured that the free energy of Ising spin glass models with the Kac potential in the non-additive limit, such as the power-law potential in the non-additive regime, is exactly equal to that of the Sherrington-Kirkpatrick model in the thermodynamic limit. 
In this study, we prove that his conjecture is true on the Nishimori line at any temperature in any dimension.
One of the key ingredients of the proof is the use of the Gibbs-Bogoliubov inequality on the Nishimori line.
We also consider the case in which the probability distribution of the interaction is symmetric, where his conjecture is true at any temperature in one dimension but is an open problem in the low-temperature regime in two or more dimensions.

\end{abstract}
\date{\today}
\maketitle

\section{Introduction}
In general, the analysis of finite-dimensional models is a challenging problem in statistical mechanics, and mean-field theory is the first starting point for theoretical analysis.
The mean-field theory is not necessarily simple in itself, and in spin glass models, the concept of replica symmetry breaking~\cite{Parisi} is a highly nontrivial phenomenon.

How valid is the picture of the mean-field theory for finite dimensional systems? 
From this perspective, the relation between the Kac potential in the non-additive limit in finite dimensions and the corresponding mean-field model has been studied in ferromagnetic models~\cite{B61,L84,PA305,JP119,E62,JPA36,Mori,Mori2,Mori3}.
``Non-additive'' means that the effective range of the interaction is over the entire system, in which the system cannot be divided and its additivity is not satisfied.
A typical example of the Kac potential in the non-additive limit is a power-low potential $1/r^{\alpha}$ in the non-additive regime $0\le \alpha < d$, where $d$ is the dimension of space.
In the Kac potential in the non-additive limit, since the interaction covers the entire system, it is naturally expected that the property of the system is similar to that of the mean-field model.
Interestingly, Mori~\cite{Mori,Mori2,Mori3} proved that, for a broad class of ferromagnetic models of classical spin systems, the free energy of the Kac potential in the non-additive limit exactly coincides with that of the corresponding mean-field model at any temperature in any dimension in the thermodynamic limit, that is, the free energy equivalence holds.
This result shows that the mean-field theory is exact for the Kac potential in the non-additive limit.

Mori~\cite{Mori} also conjectured that, by applying the analysis of ferromagnetic models to spin glass models through the replica method, a similar result holds for spin glass models at any temperature in any dimension.
The conjecture is fascinating because it suggests that the RSB picture of the mean field model is valid for long-range interacting systems.
While the replica method is not mathematically rigorous, various numerical calculations in spin glass models with the power-law potential in the non-additive regime~\cite{WY,BWM,MG,SYM} strongly suggest that the conjecture holds in at least one dimension, and we naturally expect it to hold in two or more dimensions.

In the present study, we focus on the Ising spin glass model and show that the conjecture is partially true.
First, we restrict ourselves to the Nishimori line~\cite{Nishimori,Nishimori2} and show that the free energy of the Kac potential in the non-additive limit is exactly equal to that of the Sherrington-Kirkpatrick (SK) model at any temperature in any dimension. Thus, the conjecture is true on the Nishimori line.
Next, we consider the case in which the probability distribution of the interaction is symmetric.
We prove the free energy equivalence only in the high-temperature regime in any dimension.
Furthermore, we prove that the free energy equivalence holds in one dimension at any temperature for any convex potential, such as the power-law potential in the non-additive regime, which is consistent with numerical simulations for Ising spin glass models in one dimension~\cite{WY,MG,SYM}.
Unfortunately, analysis in more than two dimensions in the low-temperature regime has not been successful.

The organization of this paper is as follows.
In Sec. II, we define the model and introduce the Kac potential in the non-additive limit and the van der Waals limit as long-range interactions.
Section III is devoted to explaining coarse graining for long-range interacting models~\cite{JP119,Mori}.
In Sec. IV, we obtain complete results on the Nishimori line.
In Sec. V, we give limited results  to the case where the distribution function of the interactions is symmetric.
Finally, our discussion is presented in Sec. VI.

\section{Model}
The Ising spin glass model with the Kac potential is defined on the $d$-dimensional lattice $Z^d$, with Ising spin variable $\sigma_i=\pm1$ and $i\in Z^d$.
Given a finite hypercube $\Lambda$ of side $L$, the Hamiltonian is given by
\be
{\mathcal{H}}_L^{(\gamma)}=-\sum_{i,j\in \Lambda}^{} \sqrt{ \frac{K(\vect{r} _{i,j}, \gamma)}{2}} \,J_{i,j}^{(\gamma)} \sigma_{i}\,\sigma_{j} \label{eqn:HD},
\ee
where  $J_{i,j}^{(\gamma)}$ are i.i.d. random variables that follow Gaussian distributions $N\qty(J_{i,j,0}^{(\gamma)},1)$, $K(\vect{r},\gamma)$ represents the interaction potential whose range is long in the sense as defined below, and the variable $\vect{r} _{i,j}$ is for the relative position of site $i$ and site $j$.
Periodic boundary conditions are imposed.

In the present paper, we consider the following two cases of $J_{i,j,0}^{(\gamma)}$: (i) the Nishimori line, $J_{i,j,0}^{(\gamma)}=\beta \sqrt{ K(\vect{r} _{i,j}, \gamma)/2}$, where $\beta$ is the inverse temperature, and (ii) the symmetric case, $J_{i,j,0}^{(\gamma)}=0$.

In order to consider the thermodynamic limit, the potential $K(\vect{r},\gamma)$ satisfies the normalized condition
\be
\sum_{i=1}^{L^d}K(\vect{r}_i,\gamma)=1, \label{normalization}
\ee
which is called the Kac's prescription.

As long-range interactions, we consider the Kac potential
\begin{equation}
K(\vect{r},\gamma)\propto \gamma^d\phi(\gamma\vect{r})\ge0\label{eqn:dfH},
\end{equation}
where $\gamma>0$.
The function $\phi(\vect{r})$ is supposed to satisfy the following conditions,
\begin{equation}
|\phi(\vect{r})|<\psi(r),\quad |\nabla\phi(\vect{r})|<\frac{\mathit{d}}{\mathit{d}r}\psi(r)\quad(\forall \vect{r},|\vect{r}|=r), \label{Kac-condition}
\end{equation}
where $\psi({r})$ is twice-differentiable, convex, integrable, and defined in $(0,\infty)$.
A typical example of the Kac potential is given by the exponential form, $K(\vect{r},\gamma)\propto \gamma^d \exp(-\gamma r)$~\cite{Kac}.

The parameter $\gamma$ characterizes the interaction range.
In order to treat the case where the effective range of the interaction is over the entire system, we will take the non-additive limit $\gamma\to 0$ with $\gamma L=1$.
Then the range of the interaction is proportional to the system size and the system loses its additivity.
It is also possible to consider the van der Waals limit $\gamma\to0$ after $L\to\infty$~\cite{LP}, where the range of the interaction is sufficiently small compared to the entire system and the additivity of the system is recovered.
Because the van der Waals limit corresponds to choosing the potential as the delta function $\delta(\vect{x})$ in the non-additive limit (see Appendix C in Ref. \cite{TN} for the proof),  
it is sufficient to consider the non-additive limit.
We also note that the Kac potential in the non-additive limit contains the power-law potential with non-additive regime, $K(\vect{r},\gamma)\propto L^{\alpha-d}/r^\alpha \ (0 \le \alpha <d)$,  because 
$ L^{\alpha-d}/r^\alpha= \gamma^d  \phi\qty(\gamma \vect{r})$ with $\phi(\vect{x})=\psi(x)=1/x^\alpha$ under the condition $\gamma L=1$. 
For these reasons, it is general enough to consider the Kac potential in the non-additive limit as long-range interactions.

The thermodynamic limit of the quenched free energy of the Kac potential in the non-additive limit, if it exists, is defined as
\be
f(\beta)&=&- \lim_{ \substack{ L\to\infty  \\  \gamma L=1} } \frac{1}{\beta L^d} \mathbb{E}\left[ \log Z_L^{(\gamma)} \right].
\ee
The corresponding mean-field spin glass model, that is, the SK model is given by
\be\label{eqn:HMF}
{\mathcal{H}}_L^{\mathrm{SK}}=-\frac{1}{\sqrt{2L^d}}\sum_{i,j=1,i\neq j}^{L^d}{ J_{i,j}^{}\,\sigma_i\,\sigma_j} ,
\ee
where $J_{i,j}$ are i.i.d. random variables that follow a Gaussian distribution $N\qty(J_{0} , 1)$ and $J_{0}$ takes two values: (i) Nishimori line, $J_0=\beta /\sqrt{{2L^d}}$ and (ii) the symmetric case, $J_0=0$.
It is well known that the thermodynamic limit of the quenched free energy of the SK model exists~\cite{GT2}
\be
f^{\mathrm{SK}}(\beta)&=&-\lim_{ \substack{ L\to\infty} } \frac{1}{\beta L^d} \mathbb{E}\left[ \log Z_L^{\mathrm{SK}}\right].
\ee

While we impose the normalization condition (\ref{normalization}) to consider the thermodynamic limit, the existence of the thermodynamic limit of the free energy is an open problem for the Kac potential in the non-additive limit in spin glass models.
For example, for the power-law potential in the non-additive regime $0\le \alpha <d$, the existence of the thermodynamic has been proved for only $\alpha=0$ (SK model) by the interpolating method~\cite{GT2} (for $\alpha\ge d$, the effective range of interaction is short and the existence of the thermodynamic limit is proved by the conventional way~\cite{KS}).
On the Nishimori line, we also answer this problem completely.

\section{coarse graining for long range interacting models}
This section briefly summarizes coarse graining for long-range interacting models, which plays a central role in the following sections.
This method was originally proposed for the power-law potential in one dimension~\cite{JP119} and was extended to the Kac potential in the non-additive limit~\cite{Mori}.
See Ref.~\cite{Mori} for the following derivation.

For any classical spin variable $S_i$ with a bounded value $|S_i|\le C$, we consider the Hamiltonian with the Kac potential
\be
\frac{1}{2L^d}\sum_{i,j\in\Lambda} K(\vect{r} _{i,j}, \gamma) S_i S_j .
\ee
Let us divide the box $\Lambda$ of volume $L^d$ into sub hypercubes $B_p$ of volume $l^d$, $p=1,2,\cdots, L^d/l^d$. 
The important point is taking the thermodynamic limit in the following order
\be
1\ll  l &\ll& L ,
\\
\gamma L&=&1 .
\ee
We introduce the local coarse graining variable $m_p$ as
\be
m_p &=&\frac{1}{l^d}\sum_{i\in B_p} S_i .
\ee
Under the conditions (\ref{eqn:dfH}) and (\ref{Kac-condition}), the following evaluation holds
\be
\frac{1}{2L^d}\sum_{i,j\in\Lambda} K(\vect{r} _{i,j}, \gamma) S_i S_j &=& \frac{1}{2L^d} \sum_{p,q=1}^{L^d/l^d} U_{pq} m_p m_q +g_1(L,l, \{ \sigma_i \}),
\ee
where $U_{pq}$ is defined as
\be
U_{pq}&=&\sum_{i\in B_p}\sum_{j\in B_q}  K(\vect{r} _{i,j}, \gamma),
\ee
and $g_1(L,l, \{ \sigma_i \})$ satisfies 
\be
\lim_{l\to\infty}\lim_{ \substack{ L\to\infty  \\  \gamma L=1} } g_1(L,l, \{ \sigma_i \})=0,
\ee
for any spin configuration.
Then, we obtain
\be
\frac{1}{2L^d}\sum_{i,j\in\Lambda} K(\vect{r} _{i,j}, \gamma) S_i S_j  &=& \frac{1}{2} \int_{C_d}d^dx  \int_{C_d}d^dy U(\vect{x}-\vect{y}) m(\vect{x})m(\vect{y})
+g_1(L,l, \{ \sigma_i \}) +\frac{1}{L^d} o\qty(L^d) , \label{conti-long}
\ee
where $C_d$ is a $d$-dimensional unit cube and $U(\vect{x})$ is defined as
\be
U(\vect{x})&=&\lim_{ \substack{ L\to\infty  \\  \gamma L=1} } L^d K(L\vect{x}).
\ee
The normalized condition corresponding to Eq. (\ref{normalization}) is given by
\be
\int_{C_d}d^dx U(\vect{x})&=&1 . \label{conti-normalization}
\ee
The continuous expression (\ref{conti-long}) is crucial in the following analysis.

\section{Complete results on Nishimori line}
In this section, we focus on the case of  the Nishimori line, $J_{i,j,0}^{(\gamma)}=\beta \sqrt{K(\vect{r} _{i,j}, \gamma) /2}$, and prove that the quenched free energy of the Kac potential in the non-additive limit exactly coincides with that of the SK model at any temperature in any dimension.
Thus, the conjecture by Mori~\cite{Mori} is true on the Nishimori line.

\subsection{Results}
Our first result is as follows.
\begin{lemma}\label{lemma1}
On the Nishimori line, the thermodynamic limit of the quenched free energy of the Kac potential in the non-additive limit, if it exists, satisfies
\be
f(\beta) &\ge& f^{\mathrm{SK}}(\beta),
\ee
at any temperature in any dimension.
\end{lemma}
The proof of Lemma \ref{lemma1} is achieved by combining the method in Sec. III with the interpolating method for Ising spin glass models with the Kac potential in the van der Waals limit~\cite{FT}.

Our second result is the inverse inequality.
\begin{lemma}\label{lemma2}
On the Nishimori line, the thermodynamic limit of the quenched free energy of the Kac potential in the non-additive limit, if it exists, satisfies,
\be
f(\beta) &\le& f^{\mathrm{SK}}(\beta), \label{NL-2nd-ineq}
\ee
at any temperature in any dimension.
\end{lemma}
The proof of Lemma \ref{lemma2} is quite different from Lemma \ref{lemma1} and is based on the Gibbs-Bogoliubov inequality on the Nishimori line~\cite{OO}.

From Lemma \ref{lemma1}, \ref{lemma2}, and the existence of the thermodynamic limit for the SK model, the squeeze theorem leads to the following result.
\begin{theorem}
On the Nishimori line, the thermodynamic limit of the quenched free energy of the Kac potential in the non-additive limit exists and satisfies
\be
f(\beta) &=& f^{\mathrm{SK}}(\beta),
\ee
at any temperature in any dimension.
\end{theorem}

\subsection{proof of Lemma \ref{lemma1}}
The idea~\cite{FT} is to interpolate between the Kac potential in a volume $|\Lambda|=L^d$ and a system constructed from a collection of many independent SK subsystem of volume $l^d$.
We devide the box $\Lambda$ into sub hypercubes $B_p$ of volume $l^d$, $p=1,2,\cdots, L^d/l^d$, and introduce the interpolating pressure function on the Nishimori line
\be
&&A_L(t)
\equiv\left(\prod_{i,j\in \Lambda} \int DJ_{i,j}^{(\gamma)} \right) \left(\prod_{p=1}^{L^d/l^d}\prod_{i,j\in B_p} \int DJ_{i,j}\right) 
\no\\
&&\log \Tr \exp(  \sum_{i,j\in\Lambda} \qty(\sqrt{tx_{i,j}^{(\gamma)}}J_{i,j}^{(\gamma)} +tx_{i,j}^{(\gamma)} )\sigma_i \sigma_j + \sum_{p}^{} \sum_{i,j\in B_p} \qty(\sqrt{(1-t)x_{i,j}}J_{i,j} +(1-t)x_{i,j} )\sigma_i \sigma_j),
\ee
where $\int Dz =\int_{-\infty}^\infty  dz\exp(-z^2/2)/\sqrt{2\pi}$, $x_{i,j}=\beta^2 /(2l^d)$ and $x_{i,j}^{(\gamma)}=\beta^2 K(\vect{r} _{i,j}, \gamma)/2$.
From the definition, the following relations hold
\be
\lim_{l\to\infty} \lim_{\substack{ L\to\infty  \\  \gamma L=1}}  \frac{1}{L^d}A_L(0)&=&\lim_{l\to\infty}\frac{1}{l^d}\mathbb{E}\left[ \log Z_l^{\mathrm{SK}}\right] = -\beta f^{\mathrm{SK}}(\beta),
\\
\lim_{l\to\infty} \lim_{\substack{ L\to\infty  \\  \gamma L=1}}  \frac{1}{L^d}A_L(1)&=& \lim_{\substack{ L\to\infty  \\  \gamma L=1}}  \mathbb{E}\left[ \log Z_L^{(\gamma)} \right]=-\beta f(\beta),
\ee
on the Nishimori line.
In the following, we will show  
\be
\dv{}{t} \frac{1}{L^d}A_L(t) \le   \frac{1}{L^d} o(L^d).
\ee
By integration by parts, we find
\be
\dv{}{t} \frac{1}{L^d}A_L(t)&=&\frac{1}{2L^d}\mathbb{E}\left[\sum_{i,j\in\Lambda}x_{i,j}^{(\gamma)}(1+\langle \sigma_i \sigma_j\rangle) -\sum_{p}^{} \sum_{i,j\in B_p}x_{i,j}(1+\langle \sigma_i \sigma_j\rangle) \right] \label{1st-deff}
\ee
where we used the identity $\mathbb{E}[\langle \sigma_i\sigma_j\rangle] = \mathbb{E}[\langle \sigma_i\sigma_j\rangle^2]$ on the Nishimori line.
From the normalization condition  (\ref{normalization}), we find 
\be
\lim_{l\to\infty} \lim_{\substack{ L\to\infty  \\  \gamma L=1}}  \frac{1}{2L^d}\qty(\sum_{i,j\in\Lambda}x_{i,j}^{(\gamma)} -\sum_{p}^{} \sum_{i,j\in B_p}x_{i,j})=0 \label{const-0}.
\ee
From Eq. (\ref{conti-long}), we find
\be
\frac{1}{2L^d}\sum_{i,j\in\Lambda} K(\vect{r} _{i,j}, \gamma) \sigma_i \sigma_j &=& \frac{1}{2} \int_{C_d}d^dx  \int_{C_d}d^dy U(\vect{x}-\vect{y}) m(\vect{x})m(\vect{y})
+g_1(L,l, \{ \sigma_i \}) +\frac{1}{L^d} o\qty(L^d) \label{conti-long-NL} .
\ee
In addition, we immediately obtain
\be
\frac{1}{2L^d} \frac{1}{l^d}\sum_{p}^{} \sum_{i,j\in B_p} \sigma_i \sigma_j&=&\frac{1}{2} \int_{C_d}d^dx m^2(\vect{x}) +\frac{1}{L^d} o(L^d). \label{conti-mean}
\ee
From Eqs (\ref{1st-deff}), (\ref{const-0}), (\ref{conti-long-NL}) and (\ref{conti-mean}), we find
\be
&&\dv{}{t} \frac{1}{L^d}A_L(t)
\no\\
&=&\frac{1}{4}\mathbb{E}\left[ \left\langle \int_{C_d}d^dx  \int_{C_d}d^dy U(\vect{x}-\vect{y}) m(\vect{x})m(\vect{y})- \int_{C_d}d^dx m^2(\vect{x}) +g_1(L,l, \{ \sigma_i \}) +\frac{1}{L^d} o(L^d)\right\rangle \right]
\no\\
&\le&\frac{1}{4}\mathbb{E}\left[ \left\langle \int_{C_d}d^dx  \int_{C_d}d^dy U(\vect{x}-\vect{y})\frac{1}{2}( m^2(\vect{x})+ m^2(\vect{y}))- \int_{C_d}d^dx m^2(\vect{x}) +g_1(L,l, \{ \sigma_i \}) +\frac{1}{L^d} o(L^d)\right\rangle \right]
\no\\
&=&\frac{1}{4}\mathbb{E}\left[ \left\langle g_1(L,l, \{ \sigma_i \})+\frac{1}{L^d} o(L^d)\right\rangle \right],
\ee
where we used  $m(\vect{x})^2+m(\vect{y})^2\ge 2 m(\vect{x}) m(\vect{y})$ and the normalized condition (\ref{conti-normalization}).
Thus, by integrating $t$ from 0 to 1 and taking the non-additive limit $L\to\infty$, $l\to\infty$ with $l/L\to0$, we arrive at
\be
&&\lim_{l\to\infty} \lim_{ \substack{ L\to\infty  \\  \gamma L=1} }  \frac{1}{L^d}\mathbb{E}\left[ \log Z_L^{(\gamma)} \right] - \lim_{l\to\infty} \frac{1}{l^d} \mathbb{E}\left[ \log Z_l^{\mathrm{SK}}\right]
\no\\
&\le& \lim_{l\to\infty} \lim_{ \substack{ L\to\infty  \\  \gamma L=1} }  \int_0^1 dt \frac{1}{4}\mathbb{E}\left[ \left\langle g_1(L,l, \{ \sigma_i \})+\frac{1}{L^d} o(L^d)\right\rangle \right]
\no\\
&=& \lim_{l\to\infty} \lim_{ \substack{ L\to\infty  \\  \gamma L=1} }   \frac{1}{4}\mathbb{E}\left[ \left\langle g_1(L,l, \{ \sigma_i \})+ \frac{1}{L^d} o(L^d)\right\rangle \right]
\no\\
&=&0,
\ee
which is a proof of Lemma \ref{lemma1}.

\subsection{proof of Lemma \ref{lemma2}}
We represent $\mathbb{E}\left[ \log Z_L^{(\gamma)} \right]$ as 
\be
\mathbb{E}\left[ \log Z_L^{(\gamma)} \right]&=& \left(\prod_{i,j=1,i\neq j}^{L^d} \int DJ_{i,j}^{(\gamma)} \right) 
\log \Tr \exp(  \sum_{i,j=1,i\neq j}^{L^d} \qty(\sqrt{x_{i,j}^{(\gamma)} }J_{i,j}^{(\gamma)} +x_{i,j}^{(\gamma)} )\sigma_i \sigma_j),
\ee
with $x_{i,j}^{(\gamma)}=\beta^2 K(\vect{r} _{i,j}, \gamma)/2$.
Here, we introduce the random field model on the Nishimori line
\be
\mathbb{E}\left[ \log Z_L^{RF} \right]&=& \left(\prod_{i=1}^{L^d} \int Dh_i \right) \log \Tr \exp(  \sum_{i=1}^{L^d} \qty(\sqrt{x_{i}}h_{i}+x_{i} )\sigma_i ),
\ee
with $x_i =\beta^2 q$.
Then, from the Gibbs-Bogoliubov inequality on the Nishimori line~\cite{OO}, we obtain
\be
\mathbb{E}\left[ \log Z_L^{(\gamma)} \right]
&\ge& \mathbb{E}\left[ \log Z_L^{RF} \right]  
+  \mathbb{E}\left[\frac{1}{2}\sum_{i,j=1,i\neq j}^{L^d} x_{i,j}^{(\gamma)} \qty(1+\langle \sigma_i \sigma_j \rangle_{RF} )  -\frac{1}{2}\sum_{i=1}^{L^d} x_{i} \qty(1+ \langle \sigma_i \rangle_{RF} )\right]
\no\\
&=&L^d \int Dy \log2\cosh\left(  \beta\sqrt{q}y+\beta^2q \right)
+  \frac{1}{2}\sum_{i,j=1,i\neq j}^{L^d} x_{i,j}^{(\gamma)} \qty(1+\qty(\int Dy \tanh\qty(  \beta \sqrt{q}y+\beta^2q ))^2 )
\no\\
&&  -\frac{1}{2}\sum_{i=1}^{L^d} x_{i} \qty(1+ \int Dy \tanh\qty(  \beta \sqrt{q}y+\beta^2q ) ),
\ee
where $\langle\cdots \rangle_{RF}$ is the thermal average with respect to the random field.
Using the normalized condition (\ref{normalization}) and by maximizing the above right-hand side with respect to $q$, we arrive at
\be
\lim_{\substack{ L\to\infty  \\  \gamma L=1}} \frac{1}{L^d}\mathbb{E}\left[ \log Z_L^{(\gamma)} \right]
&\ge& \int Dy \log2\cosh\left(  \beta\sqrt{q}y+\beta^2q \right)
+\frac{\beta^2}{4}(1-q)^2- \frac{\beta^2q^2}{2} ,
\\
q&=&\int Dy \tanh\qty(  \beta \sqrt{q}y+\beta^2q ),
\ee
in the thermodynamic limit.
Note that the above right-hand side coincides with 
\be
\lim_{ \substack{ L\to\infty} } \frac{1}{L^d} \mathbb{E}\left[ \log Z_L^{\mathrm{SK}}\right],
\ee 
on the Nishimori line~\cite{KM}.
Thus, we arrive at 
\be
\lim_{\substack{ L\to\infty  \\  \gamma L=1}} \frac{1}{L^d}\mathbb{E}\left[ \log Z_L^{(\gamma)} \right]
\ge
\lim_{ \substack{ L\to\infty} } \frac{1}{L^d} \mathbb{E}\left[ \log Z_L^{\mathrm{SK}}\right],
\ee
which is a proof of Lemma \ref{lemma2}.

\section{limited results for symmetric distribution}
In this section, we focus on the case of the symmetric probability distribution, $J_{i,j,0}^{(\gamma)}=0$, and obtain limited results compared to that of the Nishimori line.
The results in this section extend previous studies~\cite{FT,GT} on the Kac potential in the van der Waals limit to the non-additive limit.

Besides, when we consider the van der Waals limit instead of the non-additive limit in our setting, we prove that the quenched free energy of the Kac potential is exactly equal to that of the SK model at any temperature in any dimension without any assumption. 
This result had already been obtained in Ref.~\cite{FT,GT} under the assumption that all the Fourier modes of the Kac potential are non-negative.
We show that its assumption is always satisfied for the Kac potential in the van der Waals limit in any dimension.

\subsection{Results}
Our first result is as follows.
\begin{lemma}\label{lemma4}
For symmetric distribution function of the interaction, the thermodynamic limit of the quenched free energy of the Kac potential in the non-additive limit, if it exists, satisfies
\be
f(\beta) &\le& f^{\mathrm{SK}}(\beta), \label{SG-van-der-Waals} 
\ee
at any temperature in any dimension.
\end{lemma}
\begin{remark}
The inequality (\ref{SG-van-der-Waals}) was obtained for the Kac potential in the van der Waals limit in Ref.~\cite{FT}.
Lemma 4 generalizes it to the non-additive limit.
\end{remark}
The proof of Lemma \ref{lemma4} is almost same as that of Lemma \ref{lemma1}.
On the other hand, the inequality in the opposite direction of Lemma \ref{lemma4} cannot be obtained in the same way as in Lemma \ref{lemma2}, because there is no correspondence of the Gibbs-Bogoliubov inequality.
Instead, we use the annealed bound $\mathbb{E}\qty[\log Z] \le \log \mathbb{E}\qty[Z]$ and obtain the following result.
\begin{lemma}\label{lemma5}
For symmetric distribution function of the interaction, the thermodynamic limit of the quenched free energy of the Kac potential in the non-additive limit, if it exists, satisfies
\be
f(\beta) &\ge& f^{\mathrm{SK}}(\beta),
\ee
in the high-temperature regime $\beta<1$ in any dimension.
\end{lemma}
Compared to Lemma \ref{lemma2}, Lemma \ref{lemma5} is limited to the high-temperature regime $\beta<1$ and is not a satisfactory result.
In order to analyze at any temperature, we combine the method of the previous study on the Kac potential in the van der Waals limit in spin glass models~\cite{GT}, where the Fourier modes of the Kac potential play an important role, with the method in Sec. III.
We define the Fourier mode of $U(\vect{x})$ as
\be
U_{\svect{n}}&=&\int_{C_d}d^dx U(\vect{x}) \cos(2\pi \vect{n}\cdot\vect{x}) \quad (\forall \vect{n}\in\mathbb{Z}^d).
\ee
Then, we have:
\begin{lemma}\label{lemma6}
For symmetric distribution function of the interaction, if all the Fourier modes of $U(\vect{x})$ are non-negative, the thermodynamic limit of the quenched free energy of the Kac potential in the non-additive limit, if it exists, satisfies
\be
f(\beta) &\ge& f^{\mathrm{SK}}(\beta), \label{lemma6-eq}
\ee
at any temperature.
In particular, any convex function in one dimension satisfies $U_{n}\ge0$.

Furthermore, when we consider the Kac potential in the van der Waals limit instead of the non-additive limit, its Fourier modes are always non-negative, and thus
\be
f(\beta) &\ge& f^{\mathrm{SK}}(\beta), \label{van-der-Waals}
\ee
at any temperature in any dimension.
\end{lemma}
\begin{remark}
We note that Eq. (\ref{van-der-Waals}) had already been obtained in Ref.~\cite{GT} but it was assumed that all Fourier modes were non-negative.
Lemma \ref{lemma6} shows that its assumption is always satisfied.
\end{remark}

From Lemma \ref{lemma4}, \ref{lemma5}, \ref{lemma6}, and the existence of the thermodynamic limit for the SK model, the squeeze theorem implies:
\begin{theorem}\label{theorem-symm}
For symmetric distribution function of the interactions, the thermodynamic limit of the quenched free energy of the Kac potential in the non-additive limit exists and satisfies
\be
f(\beta) &=& f^{\mathrm{SK}}(\beta),
\ee
in the high-temperature regime $\beta<1$ in any dimension.

Besides, if all the Fourier modes are non-negative $U_{n}\ge0$, then
\be
f(\beta) &=& f^{\mathrm{SK}}(\beta), \label{main-convex-symm}
\ee
at any temperature.
In particular, any convex function in one dimension satisfies this condition.

Finally, when we consider the Kac potential in the van der Waals limit instead of the non-additive limit, we obtain $U_{n}\ge0$ and, thus, 
\be
f(\beta) &=& f^{\mathrm{SK}}(\beta),  \label{main-van-der-Waals}
\ee
at any temperature in any dimension.
\end{theorem}
\begin{remark}
The previous study~\cite{FT} obtained Eq. (\ref{main-van-der-Waals}) under the assumption that all the Fourier modes are non-negative.
Our result shows that its assumption is always satisfied in the van der Waals limit in any dimension.
We note that this assumption is not always satisfied in the non-additive limit, except for the convex Kac potential in one dimension.

On the other hand, Eq. (\ref{main-convex-symm}) holds for the power-law potential in the non-additive regime in one dimension.
This is consistent with the numerical calculations in one dimension~\cite{WY,MG}; thus, we obtain satisfactory results in one dimension.
Unfortunately, it remains an open problem in the low-temperature regime $\beta\ge1$ in more than two dimensions.
\end{remark}

\subsection{Proof of Lemma \ref{lemma4}}
We follow the same procedure as Lemma \ref{lemma1}.
We devide the box $\Lambda$ into sub hypercubes $B_p$ of volume $l^d$, $p=1,2,\cdots, L^d/l^d$, and introduce the interpolating pressure function
\be
B_L(t)
&\equiv&\left(\prod_{i,j\in \Lambda} \int DJ_{i,j}^{(\gamma)} \right) \left(\prod_{p=1}^{L^d/l^d}\prod_{i,j\in B_p} \int DJ_{i,j}\right) 
\no\\
&&\log \Tr \exp(  \sum_{i,j\in\Lambda} \sqrt{tx_{i,j}^{(\gamma)}}J_{i,j}^{(\gamma)}  \sigma_i \sigma_j + \sum_{p}^{} \sum_{i,j\in B_p} \sqrt{(1-t)x_{i,j}}J_{i,j} \sigma_i \sigma_j),
\ee
where $x_{i,j}=\beta^2 /(2l^d)$ and $x_{i,j}^{(\gamma)}=\beta^2 K(\vect{r} _{i,j}, \gamma)/2$.
We note that 
\be
\lim_{l\to\infty} \lim_{\substack{ L\to\infty  \\  \gamma L=1}}  \frac{1}{L^d}B_L(0)&=&\lim_{l\to\infty}\frac{1}{l^d}\mathbb{E}\left[ \log Z_l^{\mathrm{SK}}\right] = -\beta f^{\mathrm{SK}}(\beta),
\\
\lim_{l\to\infty} \lim_{\substack{ L\to\infty  \\  \gamma L=1}}  \frac{1}{L^d}B_L(1)&=& \lim_{\substack{ L\to\infty  \\  \gamma L=1}}  \mathbb{E}\left[ \log Z_L^{(\gamma)} \right]=-\beta f(\beta).
\ee
By integration by parts, we find
\be
\dv{}{t} \frac{1}{L^d}B_L(t)&=&\frac{1}{2L^d}\mathbb{E}\left[\sum_{i,j\in\Lambda}x_{i,j}^{(\gamma)}(1-\langle \sigma_i \sigma_j\rangle^2) -\sum_{p}^{} \sum_{i,j\in B_p}x_{i,j}(1-\langle \sigma_i \sigma_j\rangle^2) \right] \label{1st-deff-sym}
\ee
Introducing two replicas with identical quenched couplings and spin configuration $\sigma^1, \sigma^2$, we rewrite Eq. (\ref{1st-deff-sym}) as
\be
\dv{}{t} \frac{1}{L^d}B_L(t)&=&\frac{1}{2L^d}\mathbb{E}\left[\sum_{i,j\in\Lambda}x_{i,j}^{(\gamma)}\qty(1-\langle \sigma_i^1 \sigma_i^2 \sigma_j^1 \sigma_j^2\rangle) -\sum_{p}^{} \sum_{i,j\in B_p}x_{i,j}\qty(1-\langle \sigma_i^1 \sigma_i^2 \sigma_j^1 \sigma_j^2 \rangle) \right] \label{1st-deff-sym} .
\ee
Furthermore, we regard $\sigma_i^1 \sigma_i^2$ as a new spin variable $S_i$ and introduce the local coarse graining variable 
\be
q_p &=&\frac{1}{l^d}\sum_{i\in B_p} S_i .
\ee
Then, from Eq. (\ref{conti-long}) we obtain
\be
\frac{1}{2L^d}\sum_{i,j\in\Lambda} K(\vect{r} _{i,j}, \gamma) S_i S_j &=& \frac{1}{2} \int_{C_d}d^dx  \int_{C_d}d^dy U(\vect{x}-\vect{y}) q(\vect{x})q(\vect{y})
+g_2(L,l, \{ \sigma_i \}) +\frac{1}{L^d} o\qty(L^d) , 
\\
\frac{1}{2L^d} \frac{1}{l^d}\sum_{p}^{} \sum_{i,j\in B_p} S_i S_j&=&\frac{1}{2} \int_{C_d}d^dx q^2(\vect{x}) +\frac{1}{L^d} o(L^d). 
\ee
where 
\be
\lim_{l\to\infty}\lim_{ \substack{ L\to\infty  \\  \gamma L=1} } g_2(L,l, \{ \sigma_i \})=0.
\ee
Thus, we obtain 
\be
&&\dv{}{t} \frac{1}{L^d}B_L(t)
\no\\
&=&\frac{1}{4}\mathbb{E}\left[ \left\langle -\int_{C_d}d^dx  \int_{C_d}d^dy U(\vect{x}-\vect{y}) q(\vect{x})q(\vect{y}) + \int_{C_d}d^dx q^2(\vect{x}) -g_2(L,l, \{ \sigma_i \}) +\frac{1}{L^d} o(L^d)\right\rangle \right]
\no\\
&\ge&\frac{1}{4}\mathbb{E}\left[ \left\langle -\int_{C_d}d^dx  \int_{C_d}d^dy U(\vect{x}-\vect{y})\frac{1}{2}( q^2(\vect{x})+ q^2(\vect{y})) + \int_{C_d}d^dx q^2(\vect{x}) -g_2(L,l, \{ \sigma_i \}) +\frac{1}{L^d} o(L^d)\right\rangle \right]
\no\\
&=&\frac{1}{4}\mathbb{E}\left[ \left\langle -g_2(L,l, \{ \sigma_i \})+\frac{1}{L^d} o(L^d)\right\rangle \right],
\ee
where we used  $q(\vect{x})^2+q(\vect{y})^2\ge 2 q(\vect{x}) q(\vect{y})$ and the normalized condition (\ref{conti-normalization}).
Finally, by integrating $t$ from 0 to 1 and taking the non-additive limit $L\to\infty$, $l\to\infty$ with $l/L\to0$, we arrive at
\be
&&\lim_{l\to\infty} \lim_{ \substack{ L\to\infty  \\  \gamma L=1} }  \frac{1}{L^d}\mathbb{E}\left[ \log Z_L^{(\gamma)} \right] - \lim_{l\to\infty} \frac{1}{l^d} \mathbb{E}\left[ \log Z_l^{\mathrm{SK}}\right]
\no\\
&\ge& \lim_{l\to\infty} \lim_{ \substack{ L\to\infty  \\  \gamma L=1} }  \int_0^1 dt \frac{1}{4}\mathbb{E}\left[ \left\langle -g_2(L,l, \{ \sigma_i \})+\frac{1}{L^d} o(L^d)\right\rangle \right]
\no\\
&=&0,
\ee
which is a proof of Lemma \ref{lemma4}.

\subsection{Proof of Lemma \ref{lemma5}}
From the annealed bound $\mathbb{E}\qty[\log Z] \le \log \mathbb{E}\qty[Z]$, we find 
\be
\frac{1}{L^d}\mathbb{E}\left[ \log Z_L^{(\gamma)} \right]&\le&
\frac{1}{L^d}\log\left(  \Tr e^{  \sum_{i,j=1,i\neq j}^{L^d} \frac{1}{2} x_{i,j}^{(\gamma)}    }\right)
= \log2+\frac{1}{L^d} \sum_{i,j=1,i\neq j}^{L^d} \frac{1}{2} x_{i,j}^{(\gamma)}   
\ee
Thus we obtain
\be
\lim_{\substack{ L\to\infty  \\  \gamma L=1}} \frac{1}{L^d}\mathbb{E}\left[ \log Z_L^{(\gamma)} \right]
&\le& \log2 +\frac{\beta^2}{4}.
\ee
The above right-hand side coincides with  
\be
\lim_{ \substack{ L\to\infty} } \frac{1}{L^d} \mathbb{E}\left[ \log Z_L^{\mathrm{SK}}\right],
\ee 
in the high-temperature regime $\beta<1$, which is a proof of Lemma \ref{lemma5}.

\subsection{Proof of Lemma \ref{lemma6}}
The idea~\cite{GT} is to interpolate between the Kac potential and the SK model in a volume $|\Lambda|=L^d$.
We introduce the interpolating pressure function
\be
C_L(t)
&\equiv&\left(\prod_{i,j\in \Lambda} \int DJ_{i,j}^{(\gamma)} \int DJ_{i,j} \right) 
\no\\
&&\log \Tr \exp(  \sum_{i,j\in\Lambda} \sqrt{tx_{i,j}^{(\gamma)}}J_{i,j}^{(\gamma)}  \sigma_i \sigma_j +  \sum_{i,j\in \Lambda} \sqrt{(1-t)x_{i,j}}J_{i,j} \sigma_i \sigma_j),
\ee
where $x_{i,j}=\beta^2 /(2L^d)$ and $x_{i,j}^{(\gamma)}=\beta^2 K(\vect{r} _{i,j}, \gamma)/2$.
Using integration by parts, we find
\be
\dv{}{t} \frac{1}{L^d}C_L(t)&=&\frac{1}{2L^d}\mathbb{E}\left[\sum_{i,j\in\Lambda}x_{i,j}^{(\gamma)}\qty(1-\langle \sigma_i^1 \sigma_i^2 \sigma_j^1 \sigma_j^2\rangle) - \sum_{i,j\in \Lambda}x_{i,j}\qty(1-\langle \sigma_i^1 \sigma_i^2 \sigma_j^1 \sigma_j^2 \rangle) \right] \label{1st-deff-sym} .
\ee
Then, by the same procedure as Lemma \ref{lemma4}, we obtain 
\be
\frac{1}{2L^d}\sum_{i,j\in\Lambda}x_{i,j}^{(\gamma)} \sigma_i^1 \sigma_i^2 \sigma_j^1 \sigma_j^2 &=& \frac{1}{4} \int_{C_d}d^dx  \int_{C_d}d^dy U(\vect{x}-\vect{y}) q(\vect{x})q(\vect{y}) +\frac{1}{L^d} o(L^d),
\\
\frac{1}{2L^d}\sum_{i,j\in \Lambda}x_{i,j} \sigma_i^1 \sigma_i^2 \sigma_j^1 \sigma_j^2 &=&\frac{1}{4} \qty(\int_{C_d}d^dx q(\vect{x}))^2 +\frac{1}{L^d} o(L^d).
\ee
By the Fourier expansion, we obtain 
\be
\int_{C_d}d^dx  \int_{C_d}d^dy U(\vect{x}-\vect{y}) q(\vect{x})q(\vect{y}) &=&  \sum_{\svect{n}} U_{\svect{n}} |{q}_{\svect{n}}|^2
\\
\qty(\int_{C_d}d^dx   q(\vect{x}) )^2&=& |{q}_{\svect{0}}|^2,
\ee
where 
\be
{q}_{\svect{n}}&=&\int_{C_d}d^dx  q(\vect{x})\exp(2\pi i \vect{n}\cdot \vect{x}),
\\ 
U_{\svect{n}}&=&\int_{C_d}d^dx U(\vect{x}) \cos(2\pi \vect{n}\cdot\vect{x}) .
\ee
We note that 
\be
U_{\svect{0}}&=&\int_{C_d}d^dx U(\vect{x})  =1.
\ee
If all the Fourier modes are non-negative, $U_{\svect{n}}\ge0$, we find
\be
\int_{C_d}d^dx  \int_{C_d}d^dy U(\vect{x}-\vect{y}) q(\vect{x})q(\vect{y})-\qty(\int_{C_d}d^dx   q(\vect{x}) )^2
&=&\sum_{\svect{n}\neq \svect{0}} U_{\svect{n}} |{q}_{\svect{n}}|^2 \ge0,
\ee
and
\be
\dv{}{t} \frac{1}{L^d}C_L(t)
&=&\frac{1}{4}\mathbb{E}\left[\qty(\int_{C_d}d^dx q(\vect{x}))^2 -\int_{C_d}d^dx  \int_{C_d}d^dy U(\vect{x}-\vect{y}) q(\vect{x})q(\vect{y})+\frac{1}{L^d} o(L^d)\right] 
\no\\
&\le&\frac{1}{4}\mathbb{E}\left[\frac{1}{L^d} o(L^d)\right] .
\ee
Therefore, if $U_{\svect{n}}\ge0$, we obtain
\be
\lim_{\substack{ L\to\infty  \\  \gamma L=1}} \frac{1}{L^d}\mathbb{E}\left[ \log Z_L^{(\gamma)} \right] \le \lim_{ \substack{ L\to\infty} } \frac{1}{L^d} \mathbb{E}\left[ \log Z_L^{\mathrm{SK}}\right], \label{68-ineq}
\ee
which is a proof of Eq. (\ref{lemma6-eq}).

Next, we consider the case in which $U(\vect{x})$ is convex in one dimension.
Then, it is known that $U_{\svect{n}}\ge0$ for all $\vect{n}$ (see Ref. \cite{NR} for the proof).
Thus,  Eq. (\ref{68-ineq}) holds in the convex Kac potential in the non-additive limit at any temperature in one dimension.
Unfortunately, the same inequality does not always hold in more than two dimensions.

Finally, if we consider the van der Waals limit instead of the non-additive limit, the potential $U(\vect{x})$ is the delta function $\delta(\vect{x})$ in any dimension~\cite{Mori,TN}.
In the delta function potential, $U_{\svect{n}}=1$ for all $\vect{n}$.
Therefore, Eq. (\ref{68-ineq}) holds in the Kac potential in the van der Waals limit at any temperature in any dimension.

\section{Conclusions}
Mori has conjectured that, by the replica method, the free energy of the SK model exactly coincides with that of Ising spin glass models with the Kac potential in the non-additive limit at any temperature in any dimension.
We have shown that his conjecture is true on the Nishimori line.
One of the key elements of the proof is the Gibbs-Bogoliubov inequality on the Nishimori line.

Next, we have considered the case where the probability distribution of the interaction is symmetric, in which we are most interested.
If we take the van der Waals limit instead of the non-additive limit, we have proved that the free energy of the SK model exactly coincides with that of Ising spin glass models with the Kac potential at any temperature in any dimension.
This result removes the assumption used in previous studies~\cite{FT,GT} that all Fourier modes are non-negative.
On the other hand, when we take the non-additive limit, we have obtained the free energy equivalence in any dimension in the high-temperature regime $\beta<1$, or when all the Fourier modes of the Kac potential are non-negative. 
This means that the free energy equivalence always holds for the power-law potential in the non-additive regime in one dimension, which is consistent with the previous numerical calculations~\cite{WY,MG}.
However, it remains difficult to analyze the low-temperature regime in more than two dimensions.
The bottleneck stems from the absence of a counterpart of the Gibbs-Bogoliubov inequality.

In the previous studies~\cite{Mori}, Mori proved the free energy equivalence in ferromagnetic models by rigorously calculating the free energy.
Tsuda and Nishimori~\cite{TN} generalized Mori's method to the presence of a random magnetic field using the law of large numbers.
Our method, which does not calculate the free energy directly, also applies to their results. 
Then, the conventional Gibbs-Bogoliubov inequality enables us to perform the same procedure as in the present study for ferromagnetic models.

It is an important problem to extend the present study beyond Ising spin glass models. 
Numerical calculations proposed that similar results hold in vector spin glass models and dilute spin glass models with non-additive long-range interactions~\cite{WY,BWM}.

Finally, we mention the difficulty in diluted ferromagnetic models.
The free energy equivalence of diluted Ising spin glass models with the Kac potential in the van der Waals limit has already been obtained in previous studies~\cite{GT,FT2}, where the vanishing of odd order terms of the interaction is cleverly used.
In diluted ferromagnetic models, odd-order terms of the interactions do not vanish, and thus, the approach in spin glass models does not seem to be applicable.
To the best of our knowledge, the free energy equivalence of the diluted ferromagnetic models is an open problem not only in the non-additive limit but also in the van der Waals limit.

The present work was financially supported by JSPS KAKENHI Grant No. 19H01095,  20H02168 and 21K13848.



\end{document}